Title: **A new estimate for the age of highly-siderophile element retention in the lunar mantle from late accretion**

Authors: R. Brasser[1], S. J. Mojzsis[2,3], S. C. Werner[4] and O. Abramov[5]

Affiliations:

[1]Earth Life Science Institute, Tokyo Institute of Technology, Ookayama, Meguro-ku, Tokyo 152-8550, Japan

[2]Department of Geological Sciences, University of Colorado, Boulder, CO 80309, USA

[3]Origins Research Institute, Research Centre for Astronomy and Earth Sciences, H-1112 Budapest, Hungary

[4]Centre for Earth Evolution and Dynamics, University of Oslo, N-0315 Oslo, Norway

[5]Planetary Science Institute, Tucson, AZ 85719, USA

**Abstract**

Subsequent to the Moon's formation, late accretion to the terrestrial planets strongly modified the physical and chemical nature of silicate crusts and mantles. This alteration came in the form of melting through impacts, as well as the belated addition of volatiles and the highly siderophile elements (HSEs). Even though late accretion is well established as having been an important process in the evolution of the young solar system, its intensity and temporal decline remain subject to debate. Much of this deliberation hinges on what can be inferred about late accretion to the Moon from its computed mantle HSE abundances. Current debate centres on whether the lunar HSE record is representative of its whole late accretion history or alternatively that these were only retained in the mantle and crust after a particular time, and if so, when. Here we employ improved Monte Carlo impact simulations of late accretion onto the Moon and Mars and present an updated chronology based on new dynamical simulations of leftover planetesimals and the E-belt – a now-empty hypothesised inner extension of the asteroid belt (Bottke et al., 2012). We take into account the inefficient retention of colliding material. The source of impactors on both planetary bodies is assumed to be the same, hence we use constraints from both objects simultaneously. We compute the crater and basin densities on the Moon and Mars, the largest objects to strike these planets and the amount of material they accreted. Outputs are used to infer the mass in leftover planetesimals at a particular time period, which is then compared to the lunar HSE abundance. From this estimate we calculate a preferred lunar HSE retention age of ca. 4450 Ma which means that the modelled lunar mantle HSE abundances trace almost all of lunar late accretion. Based on our results, the surface ages of the lunar highlands are at least 4370 Ma. We find that the mass of leftover planetesimals with diameters $D_i<300$ km at 4500 Ma that best fits the crater chronology is approximately $2\times10^{-3}$ Earth mass ($M_E$) while the mass of the E-belt was fixed at $4.5\times10^{-4}$ $M_E$. We also find that a leftover planetesimal mass in excess of 0.01 $M_E$ results in a lunar HSE retention age younger than major episodes of lunar differentiation and crust formation, which in turn violates geochemical constraints for the timing and intensity of late accretion to the Earth (Mojzsis et al., 2019).

**Keywords**: cratering chronology; impact flux; late accretion; E-belt; terrestrial planets

# 1 Introduction

After the primary phase of planet accretion – wherein most of their masses were acquired – the terrestrial planets and other minor bodies (asteroids and moons) experienced a protracted history of a declining impactor flux termed *late accretion* (e.g. Dale et al., 2012; Day et al., 2012; Day et al., 2016; Mojzsis et al., 2019). In short, leftover planetesimals from the primary accretion epoch continued to evolve onto planet-crossing orbits and were variably absorbed by the terrestrial planets and large asteroids as late stage mass supplements (Wetherill, 1977). In the young solar system, late accretion manifested itself as an intense impact bombardment that thermally, structurally and chemically modified the solid surfaces of the inner planets after core formation and the initial separation of the silicate reservoirs (crust/mantle). Late accretion continues to the present day, albeit at a much reduced rate of infall.

Evidence for such late mass additions includes the elevated abundances of highly siderophile elements (HSEs; Re, Os, Ir, Ru, Pt, Rh, Pd, Au) in the mantles of Earth, the Moon and Mars, and asteroid Vesta (see Day et al., 2016 for a review). That the proportions are chondritic relative to each other, but greatly exceed their expected abundance after metal-silicate partitioning (e.g. Kimura et al., 1974) indicates that the HSEs were delivered after core formation and silicate differentiation as a late exogenous mass addition (Chou, 1978). Although the details are debated, such a late mass addition appears to be required to explain all the data. Measured abundances of these metals in terrestrial, lunar and martian samples point to a 0.7 wt.% (Day et al., 2016), 0.025 wt.% (Day & Walker, 2015) and 0.8 wt.% (Tait & Day, 2018) late accretion to these three bodies, respectively. A separate line of investigation using tungsten isotopes concludes that the measured $\varepsilon^{182}$W values in the terrestrial mantle when compared to lunar values is consistent with the Earth having acquired 0.3-0.8 wt.% of chondritic material after lunar formation (Willbold et al., 2015) that was then slowly mixed into the mantle (Frank et al., 2016). Using the same isotopic system, Touboul et al. (2015) reported that the Moon accreted up to 0.05 wt.% after its formation, somewhat more than that contingent from the lunar HSE abundances (Day et al., 2007; Day & Walker, 2015; see also Kruijer et al., 2015). An alternative interpretation of the lunar $\varepsilon^{182}$W values is that they track an early formation of the Moon (Thiemens et al., 2019). Furthermore, from partitioning experiments Brenan et al. (2019) argue that the lunar HSEs cannot be used to confidently constrain lunar late accretion (cf. Day, 2018). Last, from modelling crater formation on the Moon Richardson & Abramov (2020) conclude that the lunar craters record 0.05 wt.% late accretion to the Moon.

Bottke et al. (2010) proposed that the low HSE abundance in lunar samples is the expected result from stochastic late accretion that favoured the collision with the Earth of a few large bodies, which had remained on planet-crossing orbits after primary accretion. This stochastic accretion resulted in very little mass, mostly in the form of small planetesimals, striking the Moon. Brasser et al. (2016) built on the results of Bottke et al. (2010) and concluded that Earth's late accretion addition was primarily the result of the merger with a single lunar-sized objected dubbed *Moneta* (Genda et al., 2017). Following that analysis, Brasser & Mojzsis (2017) proposed that Mars also suffered a large collision with an approximately Ceres-sized impactor, named *Nerio* (Woo et al., 2019), that also gave rise to the formation of its satellite system (Craddock, 2011; Citron et al., 2016; Rosenblatt et al., 2016; Canup & Salmon, 2018).

The proposed colossal impact with late accretion scenarios cited above emerge from the assumption that the lunar mantle HSE abundance trace the Moon's whole history of late accretion, which in turn is used to constrain the planetesimal environment at the time of the Moon's formation and thereafter (Brasser et al., 2020). At odds with this interpretation, Zhu et al. (2019) built on work by Morbidelli et al. (2018) to reinforce the latter's argument that the lunar mantle HSEs do not trace its whole late accretion history, but instead were retained in the lunar mantle and crust only after 4350 Ma. The conclusions of Zhu et al. (2019) rest on two key premises. First, accretion of material to the Moon is only 20%-30% efficient due to the high impact speed and large masses of impactors, i.e. only 20%-30% of the impactor's mass is retained by the Moon. Zhu et al. (2019) assert a lower efficiency at earlier times, which then slowly increases as the bombardment intensity wanes. Second, they deviate from the analysis of Morbidelli et al. (2018) by assuming that the statistical analysis of lunar craters in the Neukum chronology (e.g. Neukum et al., 2001) represents both the total mass accreted by the Moon and its impact history. The measured lunar HSE abundance and the calculated accreted mass from the Neukum chronology are combined with the low retention efficiency to declare that the accreted HSEs had to have sunk to the lunar core before 4350 Ma and that the measured mantle HSE abundances are the result of the remaining of accretion after this time. The analysis by Morbidelli et al. (2018) resulted in the same age using a different chronology based on leftover planetesimals; they said an age near 4400 Ma is also possible. Morbidelli et al. (2018) conclude that the HSEs were only retained in the lunar crust and mantle after that time.

Using dynamical simulations, however, Brasser et al. (2020) showed that the rate of impacts derived from leftover planetesimals is generally inconsistent with the Neukum chronology. When the assumption of the Neukum chronology is lifted, the work by Zhu et al. (2019) removes all constraints on the amount of late accretion by the Moon over an unknown length of time. This is owing to the fact that the lunar mantle HSE abundances cannot be used to place boundaries on the amount of accretion before the closure of the system (i.e. 'lunar core closure') and, consequently, the HSE closure time is *a priori* unknown. In other words, while the system is open there is presumably no verifiable record of late accretion ever having happened.

We emphasise therefore that the lunar HSE retention age of 4350 Ma should not be considered absolute. For example, the chronology of Werner et al. (2014) yields a total mass accreted to the Moon that is roughly a factor of four lower than the Neukum chronology; assuming perfect accretion the Werner chronology delivers 0.048 wt.% to the Moon, which agrees with the 0.025 wt.% advocated by Day et al., (2007) and Day & Walker (2015) if the accretion efficiency is ~50% (Artemieva & Shuvalov, 2008). Taken together with the accretion efficiency proposed by Zhu et al. (2019), a revised (lower) amount of accretion pushes the lunar HSE retention age closer to 4500 Ma (see their Figure 3). The uncertainty over the chronology function and the resulting amount of lunar accretion motivates us to analyse how much material the Moon could accrete; we do so by combining different constraints from the Earth, Moon and Mars. The outcomes of this analysis are used to place boundaries on the mass in planetesimals and their dynamical environment at the time of the Moon's formation and shortly thereafter, and the lunar HSE retention age.

**2 Knowns and unknowns**

We begin with documenting published ages relating to core formation, differentiation, crust formation and magma ocean crystallisation on the Earth, the Moon and Mars. The purpose of this table is to show that there is a range in time wherein these planets are open systems to various radiometric sytems that facilitate HSE retention. All of these ages are older than 4300 Ma and generally younger than 4500 Ma. We follow the example of Carlson et al. (2014), but present only an abridged list of references that represent the oldest ages, rather than an exhaustive one.

**Table 1:** Collection of published ages related to core formation, differentiation and crust formation. Ages reported here are both model/intercept ages and absolute ages, but not models of ages.

| Earth | Age (Ma) | References |
|---|---|---|
| U-Pb age of silicate differentiation | 4450-4500 | Manhes et al. (1979), Albarède & Juteau (1984), Allègre et al. (2008) |
| I-Pu-Xe atmosphere retention age | 4450-4530 | Staudacher & Allègre (1982), Ozima & Podosek (1999), Mukhopadhyay (2012), Avice & Marty (2014), Caracausi et al. (2016) |
| Hf-W age of core formation | 4450-4530 | Halliday et al. (1996), Yin et al. (2002), Kleine et al. (2002, 2009) |
| Oldest-known terrestrial zircon | 4380 | Valley et al. (2014) |
| Sm-Nd silicate differentiation age | 4460-4530 | Caro et al. (2003), Boyet & Carlson (2005) |
| Lu-Hf intercept age of crust formation | 4450-4500 | Harrison et al. (2005, 2008) |
| **Moon** | | |
| Lu-Hf intercept age of crust formation | 4270-4510 | Taylor et al. (2009), Barboni et al. (2017), Maurice et al. (2020) |
| Hf-W age of lunar core formation | 4505-4530 | Jacobsen (2005), Kleine et al. (2005), Touboul et al. (2009), Thiemens et al. (2019) |
| Sm-Nd ages of silicate differentiation, magma ocean crystallisation and possible mantle overturn | 4190-4420 | Borg et al. (2011, 2015, 2019), Boyet et al. (2015), Carlson et al. (2014), Gaffney & Borg (2014), Borg et al. (2020), Maurice et al. (2020) |
| U-Pb zircon ages of magma ocean crystallisation | 4340-4370 | Taylor et al. (2009); Hopkins & Mojzsis (2015) |
| Oldest-known lunar zircon | 4417 | Nemchin et al. (2009) |
| Pb model age of lunar highlands | 4420 | Tera & Wasserburg (1974) |
| I-Pu-Xe formation age | 4463-4547 | Swindle et al. (1986) |
| **Mars** | | |
| Lu-Hf intercept age of crust formation | 4547 | Bouvier et al. (2018) |
| Hf-W age of core formation | 4560-4565 | Dauphas & Pourmand (2011) |
| Fe-Ni age of core formation | 4560-4565 | Tang & Dauphas (2014) |
| Re-Os geochron | 4550 | Brandon et al. (2000) |
| Oldest-known martian zircon | 4480 | Bouvier et al. (2018) |
| Rb-Sr isochron | 4550 | Jagoutz (1991) |
| Pb-Pb geochron | 4550 | Jagoutz (1991) |
| Sm-Nd age of silicate differentiation | 4420-4535 | Debaille et al. (2007), Nyquist et al. (2016), Borg et al. (1997, 2016) |
| I-Pu-Xe differentiation age | >4533 | Marty & Marti (2002) |

From Table 1 it appears that the Earth, Moon and Mars – as tracked by the various short- and long-lived radiogenic systems employed thus far – share a similarly dynamic history of geochemical reservoirs between 4500 Ma and 4350 Ma, during which core- and crust formation, silicate differentiation, magma ocean crystallisation and potential mantle overturn occurred. Given this evidence, we may expect that the lunar HSE retention age falls somewhere within this range.

We can see in Table 1 that for Mars there are three distinct global geochemical events that stand out: core formation near 4565 Ma, silicate differentiation and crust formation near 4550 Ma, and an event around 4480 Ma corresponding to the oldest martian zircon ages. Atmospheric closure inferred from I-Pu-Xe chronology also occurred near 4550 Ma. The expectation is therefore that the martian crust records craters after 4480 Ma and that its HSEs are retained prior to this age (e.g. Mojzsis et al., 2019).

For the Moon, the case could be made for two distinct events: the intercept ages of primary crust formation (Lu-Hf; Taylor et al., 2009; Barboni et al., 2017) and core formation (Hf-W; e.g. Jacobsen, 2005, Kleine et al., 2005, Touboul et al., 2009, Thiemens et al., 2019). These are typically older than 4500 Ma (cf. Maurice et al., 2020, who report younger ages from KREEP analysis) and could coincide with the Moon's formation. Zircon crystallisation ages and proposed timescales for silicate differentiation with potential mantle overturn and additional crust formation, happen 100-200 Myr later (cf. Maurice et al. 2020). The reason for this apparent delay is under debate, but may stem from a long-lasting (global) lunar magma ocean (Nemchin et al., 2009; Elkins-Tanton et al., 2011; Gaffney & Borg, 2014; Borg et al., 2019, Maurice et al., 2020; cf. Borg et al., 2011). Lead model ages for the lunar highlands and that of the oldest known zircon show that crust was already present, which suggests that craters on the Moon and its HSEs could be retained by 4420 Ma. An argument could also be made for later crater retention as Borg et al. (2015) presented based on Sm-Nd geochronology and a preponderance of igneous (as opposed to impact generated) lunar zircon ages clustering around 4370 Ma (Taylor et al., 2009; Hopkins and Mojzsis, 2015). Given both the large range of the Sm-Nd ages compared to the zircons and the existence of zircons older than the Sm-Nd differentiation ages, we deem it likely that crust formation and magma ocean crystallisation was already underway on the Moon before 4400 Ma.

Table 1 also shows that for the Earth, core formation appears to happen rather early, around 4530 Ma, but there is a second differentiation event recorded in U-Pb and in Sm-Nd combined with atmospheric closure in Xe isotopes that happens later, around 4480 Ma. This age also overlaps with Lu-Hf model ages of crust formation. We interpret this to mean that the terrestrial U-Pb silicate differentiation age is younger than the age of the Moon, and that a major event to affect the terrestrial geochemical reservoirs took place at that time. It can be argued that the lunar formation age and Earth's U-Pb silicate differentiation age overlap and are thus part of a single event. An explanation for a reset event of Earth's geochemical reservoirs near 4480 Ma has no simple interpretation, but might be consistent with the late impact of a lunar-sized object as advocated by Brasser et al. (2016) and Genda et al. (2017). Similarly, a global reset event at 4480 Ma on Mars may have been caused by a colossal impact (Marinova et al., 2008; Craddock, 2011; Citron et al., 2015; Brasser & Mojzsis, 2017; Bouvier et al., 2018; Woo et al., 2019). Alternatively, the 4480 Ma 'event' on Earth could record the Moon's formation.

Our interpretation of the data, however, is that lunar formation occurred at or before approximately 4510 Ma with primary crust formation commencing shortly thereafter. The lunar system closed to I-Pu-Xe (i.e., Xe outgassing) as late as 4450 Ma, began to preserve daughter-to-parent ratios in the U-Pb system at 4420 Ma, and is recorded in Sm-Nd geochronology near 4370 Ma. If this interpretation is correct, and depending on how the data are viewed, we would expect the lunar crust to begin to preserve craters from either 4420 Ma or 4370 Ma. The martian crust and mantle may have experienced partial resetting of their radiogenic systems around 4480 Ma, which would necessarily mean that its crust retains craters after this time. For the Earth, no global events that could reset radiometric systems as, for example, in the U-Pb system (e.g. Albarède & Juteau, 1984) appear to have occurred since 4480 Ma (Mojzsis et al., 2019). Taken together, it seems that for all three sampled planetary bodies global radiogenic systems with relatively high closure temperatures are effectively closed before (perhaps, well before) about 4350-4400 Ma. In our view, this means that there were no major subsequent resetting events caused by colossal impacts after ca. 4350 Ma, and thus late-accreted HSEs to the terrestrial planets should be retained before this time.

The suite of geochronological results that converge on common ages come from different radiogenic systems with different chemical affinities and a large range of half-lives and closure temperatures; they do not leave much room for interpretation apart from their uncertainties. When we attempt to tie these ages to events, such as the planetesimal environment that existed at that time, and the ensuing lunar and martian crater chronologies, the analyses become more strongly model dependent. For this, the community has turned to HSE abundances in the mantles of the (known) sampled silicate worlds as a process tracer. We have:

i   Lunar HSE abundances infer 0.025 wt.% late accretion (Day et al., 2007; Day & Walker, 2015), but this low value could be as much as 0.05 wt.% (Touboul et al., 2015), or more if it is masked by sequestration by sulphur (Rubie et al., 2016; cf. Day, 2018), uncertainty in partition coefficients (Brenan et al., 2019), or HSE sequestration prior to a specific time (Morbidelli et al., 2018; Zhu et al., 2019);
ii  Martian HSE abundances infer 0.8 wt.% late accretion (Day et al., 2016; Tait & Day, 2018); and,
iii Terrestrial HSE abundances and tungsten isotopes infer 0.7 wt.% late accretion (Day et al., 2016; Willbold et al., 2015).

Both of the latter estimates are more robust than those of the Moon. The problem with the lunar record is that there are no known direct samples of the Moon's mantle. Estimates of the Moon's mantle HSE abundances are instead derived from melt source models for lunar basaltic rock samples.

The addition of roughly 0.7 wt.% of mass to the Earth through a proposed singular large impact reset its U-Pb and Lu-Hf isotope systems, but the proposed colossal impact on Mars appeared to not have done so. This object is thought to have been only 0.1-0.3 wt.% of Mars (Brasser & Mojzsis, 2017;

Canup & Salmon, 2018), but could have been as much as 2.8 wt.% of Mars (Marinova et al., 2008; Craddock, 2011; Citron et al., 2015). Together it appears that the mass threshold to re-equilibrate the silicate reservoirs of a terrestrial planet may lie at a collision with a ~0.5 wt.% body, assuming that the impact velocities are similar.

Under the assumption that the lunar HSEs trace the Moon's whole time span of late accretion, the above three mass addition estimates can (in theory) be used to constrain the planetesimal environment at the time of the Moon's formation, resulting in a low mass and a shallow size-frequency distribution (Bottke et al., 2010; Brasser et al., 2016, 2020; Raymond et al., 2013; cf. Schlichting et al., 2012). Next, we argue how we can constrain the late accretion history of the Moon from more than just its mantle HSE abundances (see also Morbidelli et al., 2018).

Abramov & Mojzsis (2016) show that $2\times10^{22}$ kg of late accretion to Mars (3 wt.%) would cumulatively melt all of its crust. For the Earth this value becomes at least $9\times10^{21}$ kg (0.15 wt.%) (Abramov et al., 2013; Mojzsis et al., 2019) and $9\times10^{20}$ kg for the Moon (1.2 wt.%). If we accept these estimates it means that if these bodies accreted more than 0.15 wt.%, 1.2 wt.% and 3 wt.% after ca. 4500 Ma for Earth and Moon, and 4550 Ma for Mars, we would expect that all three Lu-Hf crustal intercept ages for these worlds record reset ages from wholesale fusion of their crusts. An important observation is that the threshold value of late accretion to reset the crust is the lowest for the Earth. A planetesimal has a higher probability to collide with the Earth than with Mars or the Moon (Brasser et al., 2020), and thus the terrestrial limit on late accretion places a useful upper bound on the mass in leftover planetesimals that existed at that time. Ergo, these modelling results infer upper limits on the amount of late accretion these bodies experienced after a specific time that is independent of their mantle HSE abundances, but does depend on assumptions about crustal thickness and thermal gradient.

A more robust constraint on late accretion to the Moon and Mars may be provided by their crater densities, specifically the highest crater densities that have been measured. As we argued in Brasser et al. (2020), the highest crater densities should have been produced just after crust formation. As such, the crater densities and suitable crater scaling laws yield additional information about the amount of late accretion to the Moon and Mars. Potential crater saturation, however, means that it is possible that the Moon does not retain the full ancient crater record (Head et al., 2010). For Mars, crater saturation is affected by "process saturation" such as the fluvial and aeolian activities; these work to destroy craters on Mars faster than crater saturation can accumulate (Chapman & Jones, 1977). For the Moon the difficulty is that all reliable and generally accepted calibrated ages are 3920 Ma or younger (Morbidelli et al., 2018). Therefore, we will deviate from Brasser et al. (2020) and follow the approach of Morbidelli et al. (2018) wherein we anchor our results only to the calibrated ages of 3920 Ma and after.

When the above geochronological and crater chronological ages are used to place boundaries on the outcomes of impact- and dynamical models, constraints can also begin to be placed on the mass in leftover planetesimals at 4500 Ma. If this approach is sound, when using the cratering records of the Moon and Mars the resulting planetesimal masses should agree. According to Morbidelli et al. (2018) the lunar HSE abundance and their retention age should be approximately the same as the age of a purported lunar mantle overturn and crust formation. On the other hand, large differences between a suggested lunar mantle overturn and HSE retention would imply there are either inconsistencies in the models or in our interpretation of the geochronological ages.

In this work, we extend our analyses presented in Brasser et al. (2016), Brasser & Mojzsis (2017), Mojzsis et al. (2019) and Brasser et al. (2020) by taking the lunar HSE sequestration and accretion efficiency of Morbidelli et al. (2018) and Zhu et al. (2019) into account in our Monte Carlo simulations. From this we calculate the lunar HSE sequestration age, which (as argued above) should be close to the age of crust formation and magma ocean crystallisation. We also report an updated impact chronology on the Moon based on new dynamical simulations of leftover planetesimals and the E-belt – a hypothesised inner extension to the main asteroid belt that is now empty (Bottke et al., 2012) – which yields a rate of decline of the leftover planetesimals that is slower than what was presented in Brasser et al. (2020). As we will see below, this difference in adopted bombardment histories explains some of the differences in the conclusions between our work and Zhu et al. (2019).

**3 Monte Carlo impact experiments**

We follow Brasser et al. (2020) to compute the amount of mass that impacted the Moon and Mars using Monte Carlo impact experiments. We refer Appendix A for an in-depth description of the method. Here we only discuss the differences with the original approach.

First, we treat the mass in leftover planetesimals as a free parameter whereas in the previous study it was constrained by the lunar HSE abundances and the impact probability with the Moon from various planetesimal reservoirs. Second, in this study we calculate the contribution from both leftover planetesimals and the E-belt. The two were added in series, with the leftovers first. The mass of the E-belt was fixed at $4.5\times10^{-4}$ Earth mass ($M_E$) (Bottke et al., 2012). Third, for this study we decided to use the traditional π-crater scaling law because of its wide adoption and easy comparison with previous studies. We have also run simulations with the crater scaling law of Johnson et al. (2016). We conclude that the results agree with the analyses presented below within uncertainties.

In order to account for the inefficient accretion on the Moon we modified the code as follows. Zhu et al. (2019) claim that the typical accretion efficiency on the Moon lies in between 0.2 and 0.3. We have taken the exponential relations presented for the accretion efficiency in their Figure 1 as a function of impactor diameter and we fitted these for their various values of the impact angle. Zhu et al. (2019) state that the accretion efficiency, $f_{acc}$, is given by $f_{acc} = a \exp\left(-b\frac{D_i}{D_{Moon}}\right)$, where *a* and *b* are fitting constants. We have plotted the constants *a* and *b* as a function of impact angle $\theta$ in Figure 1 for impact speeds of 15 km s$^{-1}$ and 20 km s$^{-1}$ which together encompass the mean impact speed with the Moon. We then fitted *a* with two linear segments and log *b* with a third order polynomial for an impact speed of 15 km s$^{-1}$; the fits are marginally different for an impact speed of 20 km s$^{-1}$. For each planetesimal that collides with the Moon the impact angle is uniformly distributed in cos $\theta$. From this we compute the constants *a* and *b*, and hence the accretion efficiency. The final averaged value of the accretion efficiency for all of the Moon's late accretion history is computed as a weighted average where the weights are the masses (i.e. diameters cubed) of the impactors. For consistency we now also compute the diameter of the crater excavated by each impact keeping the $\theta$ dependence.

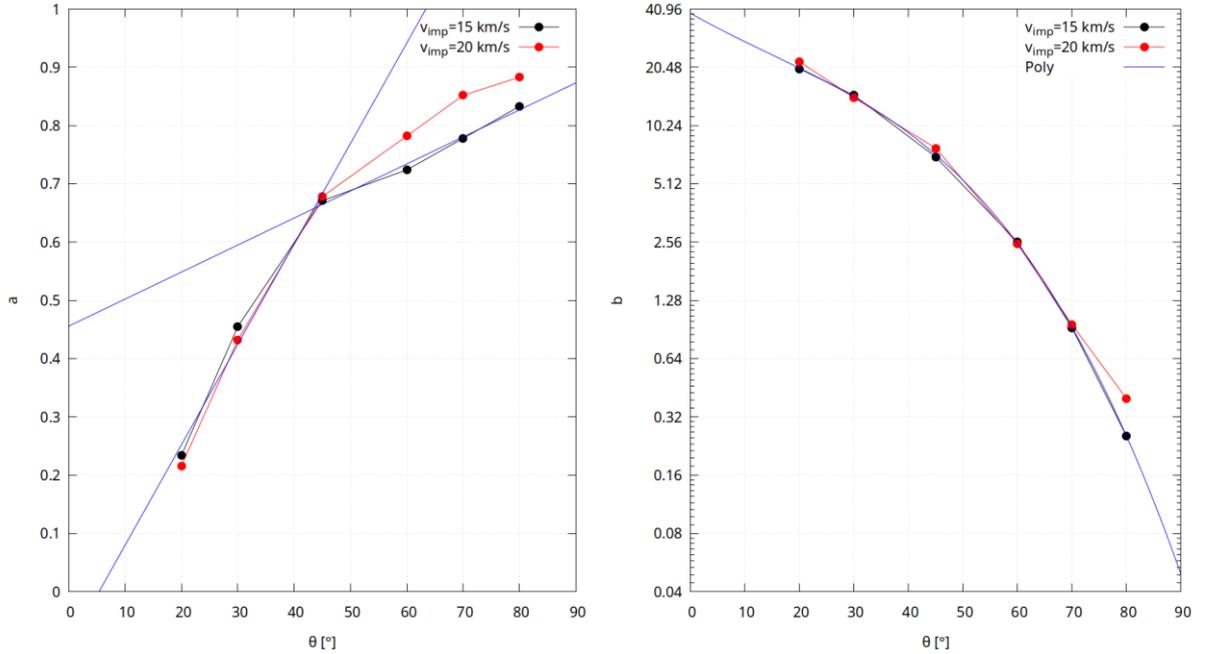

**Figure 1**. The constants *a* (left) and *b* (right) from Zhu et al. (2019) for an impact velocity on the Moon of 15 km s$^{-1}$ and 20 km s$^{-1}$. The blue lines are fits through the data for continuous values of the impact angle $\theta$.

Zhu et al. (2019) assume that planetesimals with diameter $D_i$>300 km are differentiated, the HSEs in such objects are confined to their cores and they have treated their accretion slightly differently. Here we use a similar approach: if the centre of the planetesimal with $D_i$>300 km grazes the lunar surface then it is considered to have been accreted, and if it does not then nothing was accreted.

A contour plot of the accretion efficiency as a function of impact angle and impactor diameter for an impact velocity of 15 km s$^{-1}$ is shown in Figure 2. The figure suggests typical accretion efficiencies near 0.5, whose mass-weighted average could be lowered somewhat by a few large impactors. The value of $f_{acc}$ below 0.3 suggested by Zhu et al. (2019) for the earliest epoch of late accretion is only obtained when $\theta$<20° for small impactors or when $D_i$ > 300 km and $\theta$<45°.

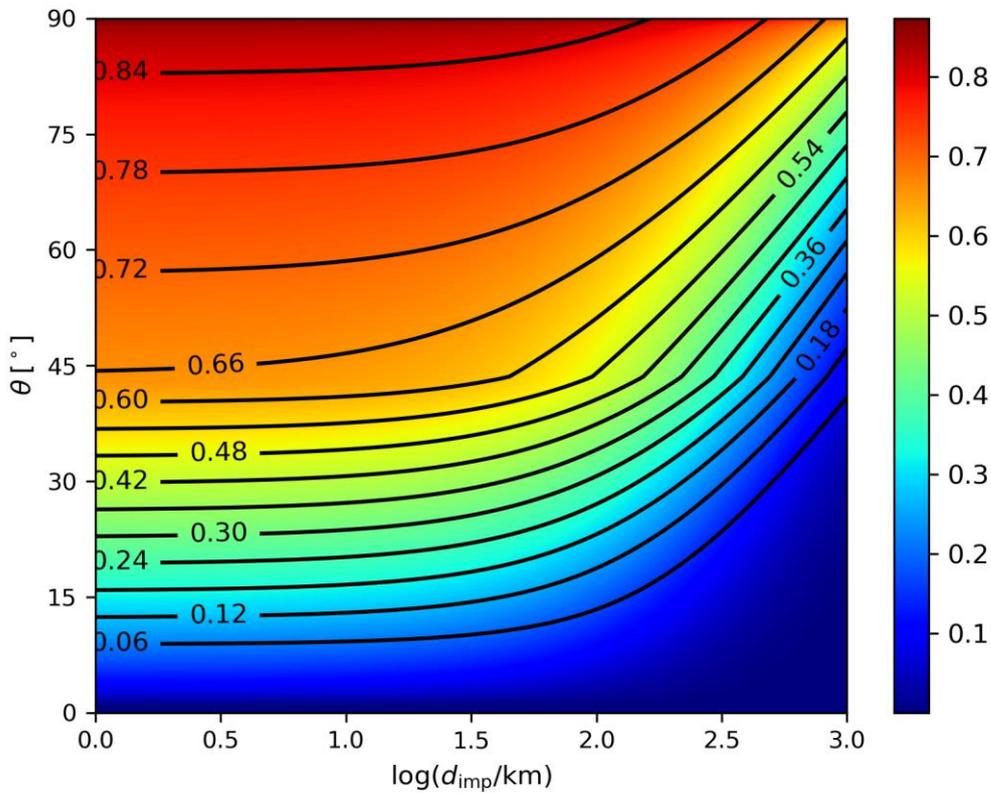

**Figure 2**. Contour plot of the accretion efficiency, $f_{acc}$, as a function of impactor diameter and impact angle. Values of $f_{acc}$ below 0.3 are only obtained when $\theta$<20° for small impactors or when $D_{imp}$>300 km and $\theta$<45°.

**4 Outcome: Planetesimal mass and HSE retention age**

*4.1 Global outcome*

The results of the Monte Carlo simulations are depicted in Figure 3. The outcomes for the Moon and Mars have been plotted together because we draw on constraints from both planets simultaneously. All quantities are plotted as a function of the logarithm of the mass in leftover planetesimals, $m_{pl}$, in units of an Earth mass. The error bars depict 2σ uncertainties that are computed from the different simulations with different random seeds. Lines show best fits through the data. We plot the mass accreted by the

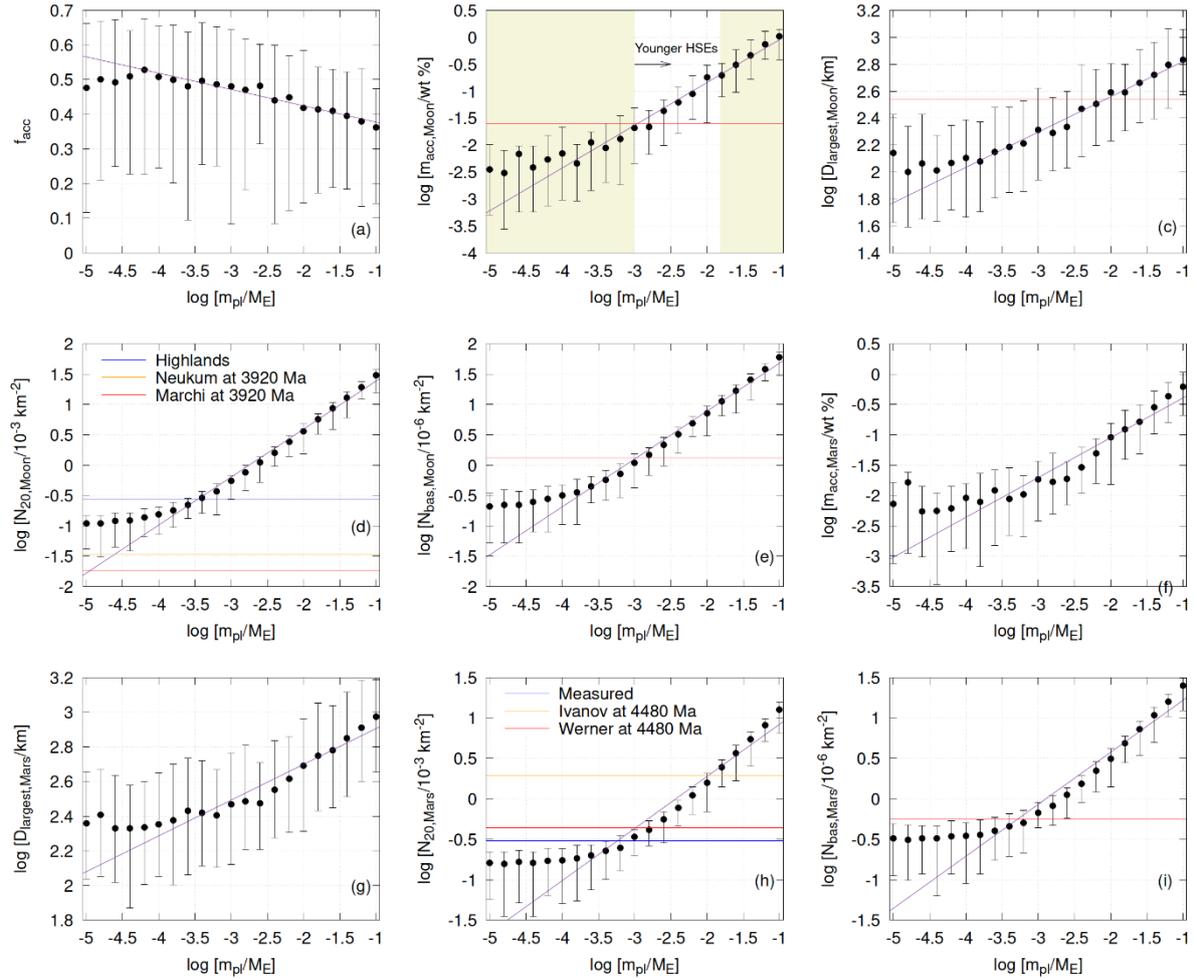

**Figure 3**. Plot of the accretion efficiency (panel a), mass accreted by the Moon (panel b), diameter of the largest impactor to strike the Moon (panel c), lunar crater density $N_{20}$ per $10^3$ km$^2$ (panel d), lunar basin density per $10^6$ km$^2$ (panel e), mass accreted by Mars (panel f), diameter of the largest impactor to strike Mars (panel g), martian crater density $N_{20}$ per $10^3$ km$^2$ (panel h) and martian basin density per $10^6$ km$^2$ (panel i) as a function of the leftover planetesimal mass. Lines show best fits through the data. Uncertainties are 2σ. Horizontal lines show observed or inferred values of accretion, crater and basin density. The shaded regions in panel 'b' are forbidden (see text for details).

Moon since 4500 Ma, roughly coincident with the earliest onset of crust formation on the Moon as estimated from other studies (Taylor et al., 2009; Barboni et al., 2017).

Panel 'a' depicts the average accretion efficiency, $f_{acc}$, of planetesimals striking the Moon. The mean accretion efficiency was weighted by the mass of the planetesimals because we are interested in total accreted mass. A simple non-weighted mean does not work because low-mass planetesimals are far more numerous than very massive ones and would skew the distribution to around 0.5. Instead, since the mantle HSE abundances are dominated by mass accretion, a mass-weighted averaging is appropriate. It is clear that $f_{acc}$ is a decreasing function of $m_{pl}$, but the average value stays above 0.4. It seldom reaches the lower value of 0.2 within uncertainty that Zhu et al. (2019) advocated early in the Moon's accretion history. The large error bars around the average value are reminiscent of a stochastic process where the outcome is heavily skewed by a few rare events.

Panel 'b' depicts the total mass accreted by the Moon, $m_{acc}$, taking the accretion efficiency into account. The horizontal line indicates the 'traditional' amount of late accretion inferred from lunar tungsten isotopes (Touboul et al., 2015) and mantle HSE abundances (Day et al., 2007; Day & Walker, 2015).

The intercept with the output from the simulations indicates an estimate of the mass in remnant leftover planetesimals that is consistent with the lunar mantle HSE budget. In other words, it shows the mass in leftover planetesimals at the time of HSE retention. The shaded regions are forbidden: on the left the mass is too low to reproduce the lunar mantle HSE abundances. The shaded region on the right is discussed in Section 4.2. The slope of the best fit line relating the accreted mass to the mass in leftover planetesimals is somewhat less than 1, so that it appears as if the Moon accretes less mass than is expected. In the Monte Carlo simulations the diameter of the planetesimals followed the size-frequency distribution of the main asteroid belt (see Appendix A). With this size-frequency distribution most of the mass is in the largest bodies. The impact probability of leftover planetesimals with the Moon is $P_{imp} = 5\times10^{-3}$ (Brasser et al., 2020), so that statistically the largest ~200 bodies will never collide with the Moon, yet they contain a sizeable fraction of the total planetesimal mass. For example, planetesimals with diameter $D_i>300$ km carry ~50% of the mass when the total mass in planetesimals $m_{pl}>10^{-3}$ $M_E$, which explains why the total accreted mass could be lower than expected. The slope between the accreted mass and the mass in leftover planetesimals is ~0.94. The flattening of the curve for low $m_{pl}$ is the result of the fixed mass of the E-belt, which dominates the mass delivery when the mass in leftover planetesimals is very low.

Panel 'c' plots the diameter of the largest object that struck the Moon in the simulations. The blue line indicates the expected maximum diameter of the planetesimal that created the South Pole-Aitken basin (Schultz & Crawford, 2011). The expected diameter of the largest planetesimal to collide with the Moon, $D_{max}$, can be computed from the fits through the data. We fit the number of generated planetesimals as a function of total planetesimal mass and multiply this with the impact probability with the Moon to obtain the total number of impacts onto the Moon. From this fit and the size-frequency distribution one can compute the expected diameter of the largest impactor, which we found to be $D_{max} = 1280(m_{pl}/M_E)^{0.32}$ km. For a leftover planetesimal mass of $10^{-2}$ $M_E$ we compute $D_{max} = 293$ km, while $D_{max}=140$ km for a leftover planetesimal mass of $10^{-3}$ $M_E$.

Panel 'd' shows the maximum density of craters with diameter $D_{cr}>20$ km, $N_{20}$. This panel assumes that all the craters are retained. The line labelled 'Measured' shows the maximum value of $N_{20}$ counted by Head et al. (2010) while 'Neukum at 3920 Ma' shows the calibrated crater density of Neukum et al. (2001), and 'Marchi at 3920 Ma' shows the calibrated data from Marchi et al. (2012). Panel 'e' depicts the calculated lunar basin density, where the horizontal line corresponds to the measured number of basins from Neumann et al. (2015).

We now move on to Mars. Panel 'f' plots the total amount of mass accreted by Mars. Panel 'g' depicts the largest planetesimal that struck Mars while the panel 'h' shows the maximum martian crater density $N_{20}$. The 'measured' line comes from Werner (2014) while the others show the calculated crater densities from the Ivanov (2001) and Werner (2019) chronologies at 4480 Ma with the crater production function of Ivanov (2001). Last, panel 'i' shows the density of martian basins where the horizontal line is the measured value from Werner (2014).

*4.2 HSE retention age*

There are several aspects of HSE retention that demand a short discussion. To begin, the lunar mantle HSEs, the lunar basins and the martian basins more or less all infer the same mass in leftover planetesimals. The mass in leftover planetesimals declines with time, so that the intercept in panel 'b' is akin to a time stamp. Essentially, the retention of the lunar mantle HSEs and the lunar and martian basins all correspond to roughly *the same point in time*. Both of the lunar and martian highlands are slightly younger. This argument was made by both Morbidelli et al. (2018) and Zhu et al. (2019), who suggested that the Moon does not retain all craters and all basins because the warm lithosphere did not support them (e.g. mechanical relaxation).

Next, the lunar mantle HSE abundances and crater record infer a leftover planetesimal mass of approximately $10^{-3}$ $M_E$. The question remains, when did this occur? The measured martian crater and densities point to a very similar amount of leftover planetesimal mass at 4480 Ma. Such a low mass in leftover planetesimals could be problematic to damp the eccentricities and inclinations of the terrestrial planets to their current values because there is little mass to provide dynamical friction. If the terrestrial planets formed very quickly (Lammer et al. 2020) this issue could become moot. However, the Moon-forming event occurred after the removal of gas from the disk and it should have caused some dynamical stirring to damp the eccentricities (Bottke et al. 2015).

We now turn to the HSE retention age. Zhu et al. (2019) use the Neukum chronology function to compute the total amount of mass that strikes the Moon and multiply this by their value of $f_{acc}$ to calculate the total mass that is retained in the lunar mantle in the form of late accretion. They compare this with the amount implied from the lunar HSE record and from there derive the HSE sequestration age. The total amount of late accretion to the Moon recorded by the HSEs is approximately $m_{HSE}$=0.025 wt.% (Day et al., 2007; Day & Walker, 2015), which corresponds to $3 \times 10^{-6}$ $M_E$. The inferred mass in leftover planetesimals consistent with the lunar HSE budget is calculated as

$$m_{\text{pl,a}} = \frac{m_{\text{HSE}}}{f_{\text{acc}} P_{\text{imp}}}. \quad (1)$$

Let us now revisit the calculation by Zhu et al. (2019). Following Morbidelli et al. (2018) we convert the crater density from the Neukum chronology at 4500 Ma to a mass colliding with the Moon, which is 0.22 wt.% or $2.64 \times 10^{-5}$ $M_E$, of which ~25% is retained (Zhu et al., 2019), i.e. $m_{acc}=1 \times 10^{-5}$ $M_E$. This is roughly a factor of three higher than the accreted mass inferred from the lunar HSE record, so the age at which the HSEs are retained is when the Neukum curve – normalised to 1 at 4500 Ma – intersects the value of $m_{HSE}/m_{acc}$ (which is approximately 0.3). This resultant age is 4325 Ma, close to the value of 4350 Ma advocated by Zhu et al. (2019). However, as we stated from the outset, this age should not be considered absolute. It will be very different when using the Werner chronology, or when relying on measured crater and basin densities, or when the accretion efficiency is different, or when the chronology function is taken to be that of leftover planetesimals (Morbidelli et al., 2018).

In a more general sense, the HSE retention age is computed as the time at which the accreted mass exceeds the lunar HSE budget. Setting $<f_{acc}>$ = 0.45 in equation (1) we calculate that the leftover planetesimal mass that satisfies the lunar mantle HSE abundances is $m_{pl,a}=1.3 \times 10^{-3}$ $M_E$. This suggests that the HSEs are retained after 4500 Ma only if $m_{pl} > 1.3 \times 10^{-3}$ $M_E$ at 4500 Ma. The question of when the HSEs are retained on the Moon boils down to the determination of the mass in leftover planetesimals at the time of the formation of the Moon. Crater densities can be converted to an accreted mass, so that the problem becomes one of matching the decline of the leftover planetesimals to the calibrated data of the lunar chronology.

This is shown in Figure 4, where we plot the decline of the E-belt (green dots), leftover planetesimals (blue dots) and the calibrated crater densities of Neukum et al. (2001) and Marchi et al. (2012) scaled from their tabulated values of $N_1$ with appropriate scaling from their respective cater production functions. The leftover planetesimal impact chronology data in the top panel come from simulations presented in Brasser et al. (2020) which were extended to 2.5 Ga. The E-belt data are from new simulations with identical initial conditions to those of Bottke et al. (2012) and Brasser et al. (2020) with more planetesimals that were run for 2 Gyr. The leftover planetesimals in the bottom panel come from new simulations run for 2 Gyr wherein planetesimals with initial perihelion distance $q$>1.5 au were also kept (about 5% of the total). The pink dashed line shows the crater density corresponding to the mass accreted to reproduce the lunar HSE abundances, the blue dashed line is the highest measured highland crater density from Head et al. (2010) and the purple line is a fit through the combined E-belt and leftover planetesimal decay.

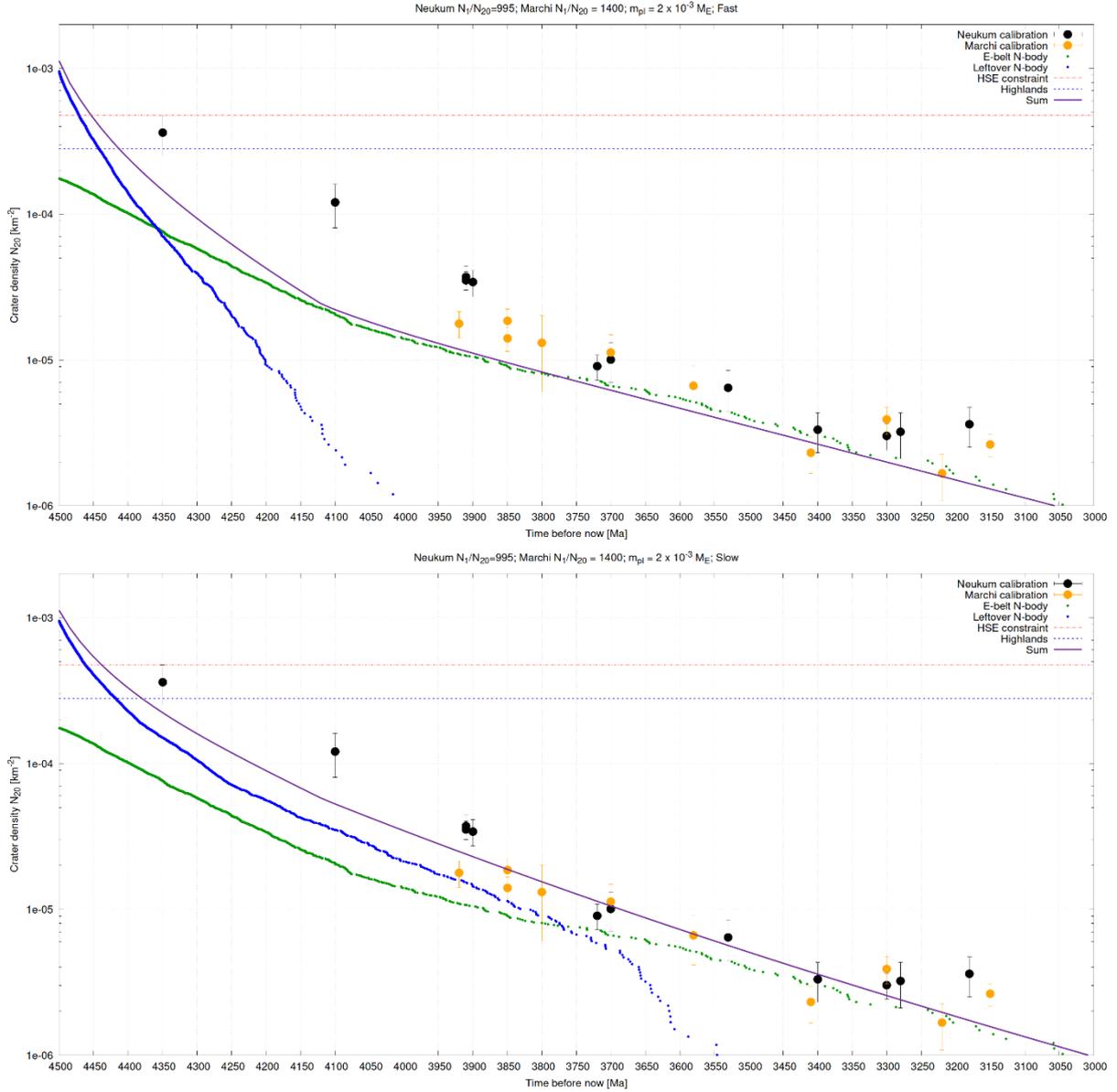

**Figure 4**. Lunar chronology based on dynamical simulations. The green dots are impacts from the E-belt, the blue dots are from leftover planetesimals. The black and orange dots are the calibrated Neukum and Marchi densities. The dashed pink line is the crater density corresponding to the mass accreted to reproduce the lunar HSE abundances, while the dashed blue line corresponds to the highland crater density from Head et al. (2010). The purple line is the chronology fitted through both components. The top panel is for the fast leftover decay presented in Brasser et al. (2020). The bottom panel is created from new dynamical simulations wherein all the leftover planetesimals were kept.

From experimentation with the bottom panel we deduce that a leftover planetesimal mass of $2\times10^{-3}$ $M_E$ produces the best fit through the calibrated data from Neukum and Marchi from 3920 Ma onwards. With this mass in leftover planetesimals the lunar HSE retention age is ca. 4450 Ma. The highland surface age is estimated to be ca. 4370 Ma, which approximately coincides with a suite of zircon and Sm-Nd differentiation ages (e.g. Borg et al., 2015; Borg et al., 2020). The leftover mass in the top panel is less well constrained due to its rapid decline; we plotted the same mass as in the bottom panel for easy comparison. This still results in an HSE retention age of ca. 4450 Ma because for the first ~100 Myr the rate of decline of the leftover planetesimals in both panels is almost the same.

We can use the two different impact chronologies in Figure 4 to constrain the maximum mass in leftover planetesimals at 4500 Ma consistent with both the geochemical and geochronological evolution of the Moon. Morbidelli et al. (2018) suggest that their calculated HSE retention age of 4350 Ma coincides with the preferred lunar crust formation age of Borg et al. (2011). The published Sm-Nd differentiation ages of the Moon, together with the oldest clustering of U-Pb zircon formation ages near 4370 Ma (e.g. Taylor et al. 2009; Hopkins & Mojzsis, 2015) represent major events in the Moon's internal evolution and crust formation (see Table 1). Thus it becomes feasible to say that 4350 Ma is the minimum age for the retention of the lunar mantle HSEs. By adding or subtracting mass to the leftover planetesimal population we report that the mass in leftover planetesimals with $D_i$<300 km that results in an HSE retention age of 4350 Ma for the fast decay is 0.01 $M_E$, while that for the slow decay is $5\times10^{-3}$ $M_E$. The first of these estimates is the same as that found by Morbidelli et al. (2018) while the second is the same as that inferred from the Neukum chronology. Thus, we argue that 0.01 $M_E$ can be viewed as a firm upper limit on the mass in leftover planetesimals with $D_i$<300 km at 4500 Ma that is independent of the terrestrial zircon record.

The age of the oldest lunar zircon is 4417 Ma (Nemchin et al., 2009) which thus records the earliest direct evidence of crust formation on the Moon. Once a crust has formed it becomes more difficult for impactors to penetrate it and deliver their HSEs to the lunar mantle (Zhu et al., 2019). Our estimated HSE retention age is younger than this.

## 5 Conclusions

Based on Monte Carlo impact experiments with the Moon and Mars, and taking into account the efficiency of lunar accretion and an updated chronology from dynamical simulations, the mass in leftover planetesimals with $D_i$<300 km at 4500 Ma that provided source materials for late accretion is approximately $2\times10^{-3}$ $M_E$ – the mass in leftover planetesimals with $D_i$<300 km at 4500 Ma is constrained to be in between $10^{-3}$ and $10^{-2}$ $M_E$. From this we deduce that the lunar mantle HSEs were very likely retained around 4450 Ma. We argue that this is the expected result because retention should likely occur before or during the onset of crust formation. Using the same chronology we suggest that the lunar highlands are at least 4370 Ma. The low leftover planetesimal mass has implications for the masses available to account for late accretionary bombardments to the inner solar system, for the interpretation of radiometric clocks for the crusts and atmospheres of the terrestrial planets (including Venus when data become available), and for the nature of the thermal environment(s) present on Earth (and Mars) at the dawn of the biosphere.

It is also clear that the calculated HSE retention age is dependent on the assumed impact chronology. The age published by Zhu et al. (2019) is a direct result of their usage of the Neukum chronology. A later assumed date of crust formation, for example near the zircon age peak of 4370 Ma (Borg et al. 2020) does not change the outcome that the HSE retention age depends on the assumed chronology model.

**Acknowledgements**

We thank two anonymous reviewers for valuable feedback. RB acknowledges financial assistance from the Japan Society for the Promotion of Science (JSPS) Shingakujutsu Kobo (JP19H05071). OA and SJM acknowledge funding from the NASA Solar System Workings Program grant 80NSSC17K0732. SCW is grateful for financial assistance from the Research Council of Norway through the Centre of Excellence funding scheme, project number 223272 (Centre for Earth Evolution and Dynamics). The Monte Carlo simulations were performed on resources provided by UNINETT Sigma2 – the National Infrastructure for High Performance Computing and Data Storage in Norway. The source code for the Monte Carlo simulations is available upon request from RB.

**Appendix A**

For the Monte Carlo code we employ the same methodology and size-frequency distribution of the planetesimals as described in Brasser et al. (2020). The size-frequency distribution (SFD) of the planetesimals is assumed to be identical to that of the present main asteroid belt. We approximate the SFD as a simple stepwise power-law distribution: for diameters 1≤D≤100 km the cumulative slope is α=2.11 (Bottke et al., 2005; Masiero et al., 2015); for larger planetesimals the cumulative slope is α=3. This size-frequency distribution is parametrised with a uniform random number $\xi$ as $D_i = \xi^{-\frac{1}{2.11}}$ for $\xi$>100$^{-2.11}$ and $D_i = 10^{0.593}\xi^{-\frac{1}{3}}$ otherwise. For each planetesimal that was generated we compute two random numbers uniformly between 0 and 1. The first random number determines impact occurs with the Moon. The second random number is used to determine an impact occurs with Mars. If either of these two random numbers is lower than the impact probability with the corresponding planet we assume that the planetesimal strikes the surface of said planet and we generate the next planetesimal. Impact probabilities for the leftover planetesimals are 0.51% for the Moon and 1.3% for Mars, and 0.25% and 2.2% for the E-belt (Brasser et al., 2020). If none of the random numbers are lower than the impact probabilities with the planets then we assume no impact occurred and we generate the next planetesimal. We sum both the total mass of all the generated planetesimals and how much mass impacts the surface of each planet. The simulation continues to generate new planetesimals until the total mass in planetesimals exceeds some pre-defined value, which ranges from $10^{-5}$ $M_E$ to 0.1 $M_E$ in logarithmic intervals of $10^{0.2}$; if the total mass in planetesimals was >25% higher than the threshold value, for example because of the sporadic generation of very large planetesimals, the output was discarded and the simulation was reset from the beginning. Once it had exhausted the mass in leftover planetesimals it carried on calculating the contribution from the E-belt. For each value of the total mass in planetesimals we run 100 individual simulations, for a total of 2100.

Planetesimals have a minimum diameter of 1 km but no maximum value. We used a bulk density of 2500 kg m$^{-3}$ for all planetesimals, which is a value designed to account for all compositions and the assumed impactor's SFD (Morbidelli et al., 2012). Smaller planetesimals will contribute towards the smaller crater population but add very little to the overall accreted mass.

The code tracks how many basins (defined as craters with diameter >300 km) as well as craters with diameter >20 km are formed on the Moon and Mars. For this study we decided to use the traditional π-scaling law because of its wide adoption and easy comparison with previous studies. Upon impact, a transient crater with diameter $D_{tr}$ is created, whose diameter scales with impactor diameter, $D_i$, as (Schmidt & Housen, 1987)

$$D_{tr} = 1.16 \left(\frac{\rho_i}{\rho_t}\right)^{1/3} D_i^{0.78} v_i^{0.44} g_t^{-0.22} (\sin\theta)^{0.44} \quad (A1),$$

where subscript $i$ stands for the impactor and $t$ for the target. Here $\rho_t$ is crustal density of the target (2700-2900 km m$^{-3}$ for the Moon and the terrestrial planets), $v_i$ is the impact speed, $g_t$ is the acceleration due to gravity of the target and $\theta$ is the impact angle. For large complex craters a transition occurs that relaxes the outer walls which further increases its diameter. For these larger craters the final diameter can be computed via (Croft, 1985)

$$D_{tr} = D_f^\eta D_{SC}^{1-\eta} \quad (A2),$$

where typically η=0.85 and where $D_{SC}$ is the crater diameter at which the transition from simple to complex craters occurs. The final crater diameter, $D_f$, scales as the impactor diameter $D_i^{0.92}$. The value of $D_{SC}$ for silicate worlds can be calculated from (Pike, 1980)

$$D_{SC} = 24 \left(\frac{1 \text{ m s}^{-1}}{g}\right) \text{ km} \quad (A3).$$

Johnson et al. (2016) used a different relation to compute the final crater diameter from the transient crater diameter (equation A2). Their treatment results in a slightly shallower dependence of the crater diameter upon impactor diameter and larger crater diameter for the same impactor diameter; the final crater diameter scales as impactor diameter $D_i^{0.88}$. Here we make use of the mean impact velocities from the simulations of Brasser et al., (2020). The lunar impact velocity distribution approximately follows an offset Rayleigh distribution $F(>v) = 1 - \exp\left[-\frac{(v-\mu)^2}{2\sigma^2}\right]$. A fit through the impact velocity data yields an offset of $\mu$=3.0 km s$^{-1}$ and a scale parameter of $\sigma$=12.9 km s$^{-1}$. The mean impact velocity with the Moon is $18.9^{+14.9}_{-12.6}$ km s$^{-1}$ (2$\sigma$). Assuming an impact angle $\theta$=45°, an impactor speed $v_i$=18 km s$^{-1}$ and impactor density $\rho_i$=2500 kg m$^{-3}$ on the Moon with a crustal density of 2700 kg m$^{-3}$ the diameter of a planetesimal excavating a 20-km crater would be $D_i$=1.43 km. Using the same scaling relations a planetesimal diameter of $D_i$=27.4 km is required for creating a basin i.e. a crater with $D_f$>300 km. For Mars the scaling parameter is $\sigma$=7.9 km s$^{-1}$ and offset $\mu$=3.0 km s$^{-1}$. The mean impact speed with Mars is $v_i$=$14.3^{+11.7}_{-9.2}$ km s$^{-1}$ so that $D_i$=1.74 km and $D_i$=33.3 km for excavating a 20-km crater and a basin respectively. With the crater scaling law of Johnson et al. (2016) assuming an impact angle $\theta$=45° the threshold planetesimal diameters for excavating 20-km craters on the Moon and Mars are 1.0 km and 1.1 km respectively. For basins the threshold planetesimal diameters are 17 km and 22 km respectively.

**Appendix B**

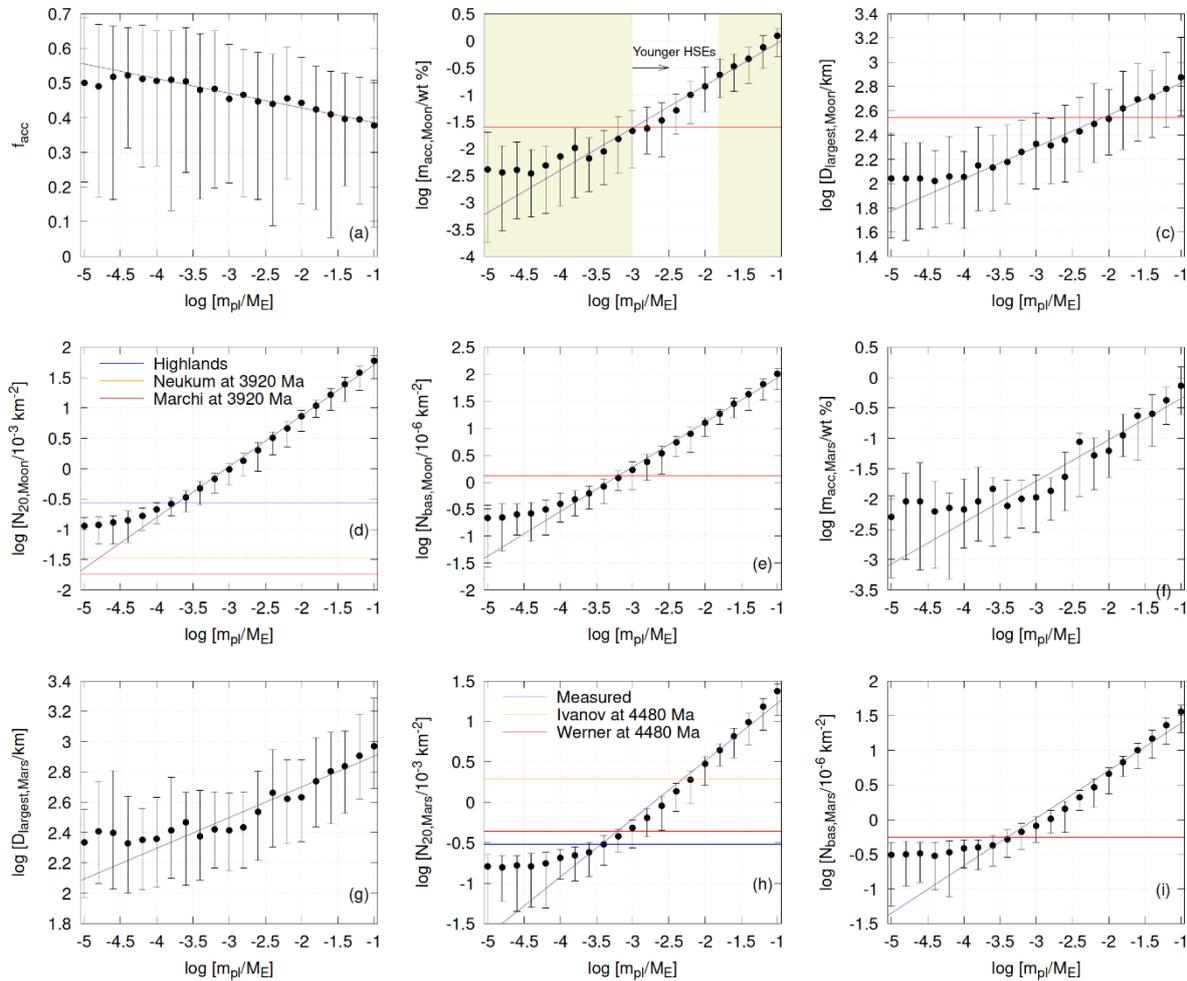

**Figure B1**. The same as Figure 3, but this uses the Johnson crater scaling law. The number density of craters is higher for a fixed planetesimal mass, but the overall results are similar to those with the previous scaling law.

In this Appendix we present the results of the Monte Carlo simulations using the crater scaling laws of Johnson et al. (2016). Figure B1 is similar to Figure 3. The Johnson crater scaling law roughly doubles the crater density for a fixed mass in planetesimals, so that the lunar and martian craters predict a lower planetesimal mass at the time that the highlands formed on both bodies. Figure B2 is similar to Figure 4. The nominal E-belt is slightly mismatched to the Neukum and Marchi calibration points: it produces too many craters to comfortably fit the calibration points. The mass in leftover planetesimals from the slow decline that matches the HSE constraint is decreased to $1\times10^{-3}$ $M_E$. The HSE retention age then becomes ca. 4480 Ma and the highlands are at least 4340 Ma.

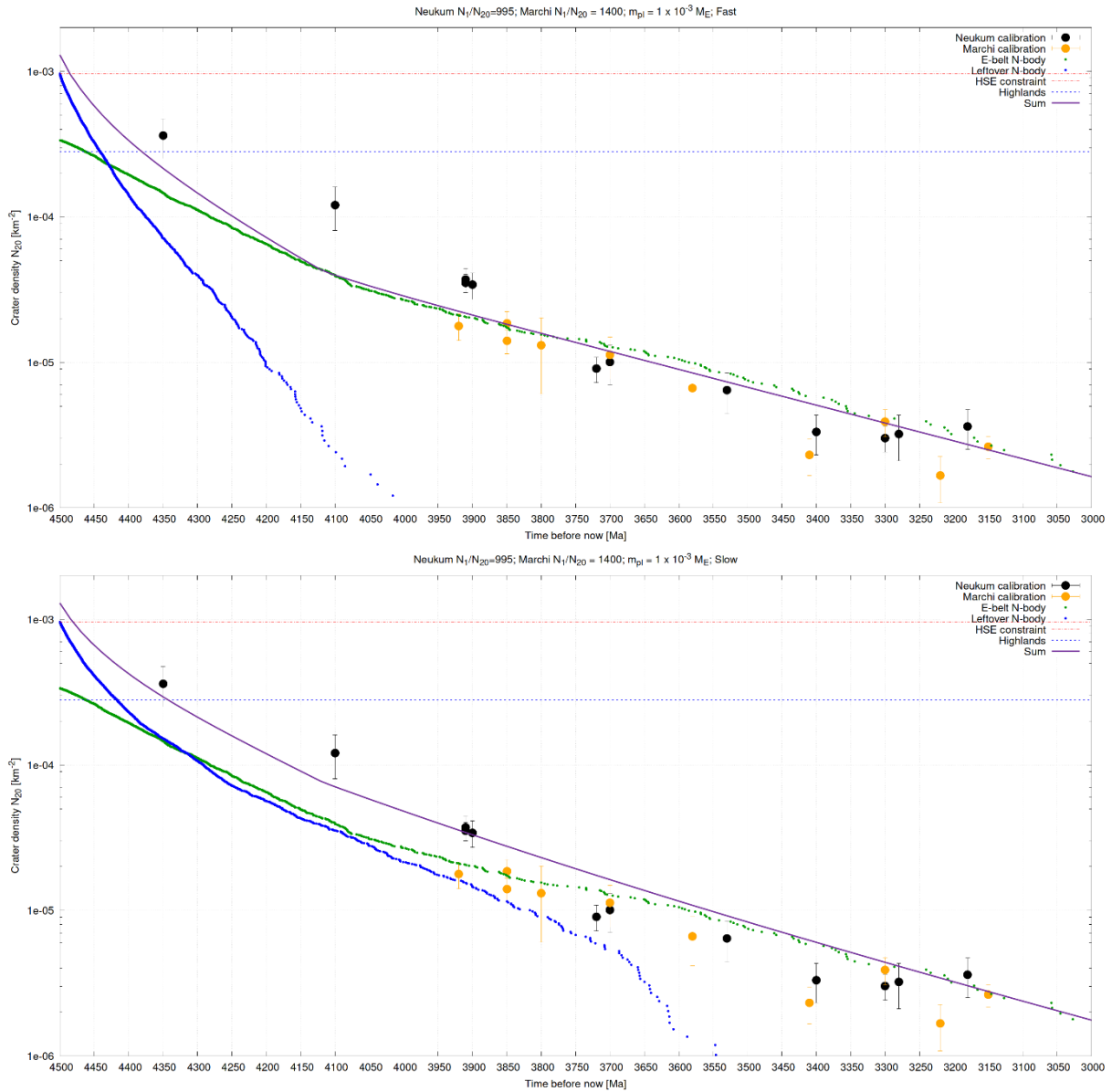

**Figure B2**. Same as Figure 4, but now using the Johnson crater scaling law. The higher crater densities cause a slight mismatch between the nominal E-belt and the calibration points from Neukum and Marchi.